\input harvmac.tex
%
\figno=0
\def\fig#1#2#3{
\par\begingroup\parindent=0pt\leftskip=1cm\rightskip=1cm\parindent=0pt
\baselineskip=11pt
\global\advance\figno by 1
\midinsert
\epsfxsize=#3
\centerline{\epsfbox{#2}}
\vskip 12pt
{\bf Fig. \the\figno:} #1\par
\endinsert\endgroup\par
}
\def\figlabel#1{\xdef#1{\the\figno}}
\def\encadremath#1{\vbox{\hrule\hbox{\vrule\kern8pt\vbox{\kern8pt
\hbox{$\displaystyle #1$}\kern8pt}
\kern8pt\vrule}\hrule}}

\overfullrule=0pt

%
\def\cqg#1#2#3{{\it Class. Quantum Grav.} {\bf #1} (#2) #3}
\def\np#1#2#3{{\it Nucl. Phys.} {\bf B#1} (#2) #3}
\def\pl#1#2#3{{\it Phys. Lett. }{\bf B#1} (#2) #3}

\def\physrev#1#2#3{{\it Phys. Rev.} {\bf D#1} (#2) #3}
\def\ap#1#2#3{{\it Ann. Phys.} {\bf #1} (#2) #3}

\font\zfont = cmss10 

\def\bigone{\hbox{1\kern -.23em {\rm l}}}
\def\ZZ{\hbox{\zfont Z\kern-.4emZ}}

\def\CN{{\cal N}}
\def\CV{\hat {\cal V}}
\def\hv{\hat v}

\def\a{\alpha}
\def\b{\beta}

\def\k{\kappa}

\def\m{\mu}
\def\n{\nu}

\def\G{\Gamma}

\def\o{\over}

\def\xih{{\hat \xi}}
\def\vh{{\hat {\cal V}}}

\Title{
{\vbox{
\rightline{\hbox{CALT-68-2379}}
\rightline{\hbox{UMD-PP-02-042}}
\rightline{\hbox{ROM2F/2002/10}}
}}}
{\vbox{\hbox{\centerline{Supersymmetry Breaking and
${\alpha'}$-Corrections}}
\hbox{\centerline{to Flux Induced Potentials }}}
}
\centerline{Katrin Becker}
\centerline{\it California Institute of Technology 452-48,
Pasadena, CA 91125}

\centerline{Melanie Becker}
\centerline{\it Department of Physics, University of Maryland,}
\centerline{\it College Park, MD 20742-4111}

\centerline{Michael Haack}
\centerline{\it Dipartimento di Fisica, Universit\`a di Roma 2, ``Tor
Vergata",}
\centerline{\it 00133 Rome, Italy}

\centerline{Jan Louis}
\centerline{\it Fachbereich Physik, Martin-Luther-Universit\"at
Halle-Wittenberg,}
\centerline{\it Friedemann-Bach-Platz 6, D-06099 Halle, Germany}

\baselineskip 18pt

\vskip 8pt

\noindent

We obtain the vacuum solutions for ${\cal M}$-theory compactified on
eight-manifolds with non-vanishing four-form flux by analyzing
the scalar potential appearing in the three-dimensional theory.
Many of these vacua are not supersymmetric and yet have a vanishing
three-dimensional cosmological constant.
We show that in the context of Type IIB
compactifications on Calabi-Yau threefolds with fluxes and external brane
sources $\alpha'$-corrections generate
a correction to the supergravity potential proportional to the
Euler number of the internal
manifold which spoils the no-scale structure appearing in the classical
potential. This indicates that $\alpha'$-corrections may indeed
lead to a stabilization of the radial modulus appearing in these
compactifications.

\Date{April 2002}

\newsec{Introduction}

Compactifications of ${\cal M}$-theory on a Calabi-Yau fourfold
result in a theory with ${\cal N}=2$ supersymmetry in three dimensions
which is roughly equivalent to
${\cal N}=1$ supersymmetry in
four dimensions. However, there is no evidence for supersymmetry
in the observed low energy spectrum, so that physics below the
compactification scale must break supersymmetry in such a way that
the cosmological constant remains extremely small.
Some time ago gaugino condensation was suggested as a possible
mechanism for hierarchical supersymmetry breaking
in compactifications of the heterotic string
\ref\drsw{M.~Dine, R.~Rohm, N.~Seiberg and E.~Witten,
``Gluino Condensation in Superstring Models'',
\pl {156} {1985} {55}.}, \ref\DerendingerKK{
J.~P.~Derendinger, L.~E.~Ibanez and H.~P.~Nilles,
``On The Low-Energy D = 4, N=1 Supergravity Theory Extracted from the D =
10, N=1 Superstring'', \pl {155} {1985} {65}.}.
Augmented with the possibility of a background flux for
the field strength of the antisymmetric tensor
\ref\RohmJV{R.~Rohm and E.~Witten,
``The Antisymmetric Tensor Field in Superstring Theory'',
\ap {170} {1986} {454}.} it was observed
that in simple models a stable vacuum with broken supersymmetry
and vanishing cosmological constant can be arranged.
However, it has been problematic to implement  a large
hierarchy of scales and a realistic pattern of soft
supersymmetry breaking terms
\ref\NillesAP{For a recent review see, for example,
H.~P.~Nilles, ``Hidden Sector Supergravity Breakdown'',
{\it Nucl.\ Phys.\ Proc.\ Suppl.\ } {\bf 101} (2001) 237, hep-ph/0106063.}.
Recently a similar scenario has been revived
in the context of warped Type IIB compactifications
on Calabi-Yau threefolds with three-form fluxes
 and localized sources
\ref\gkp{S.~B.~Giddings, S.~Kachru and J.~Polchinski,
``Hierarchies from Fluxes in String Compactifications'',
hep-th/0105097.}, \ref\KST{S.~Kachru, M.~Schulz and S.~Trivedi,
``Moduli Stabilization from Fluxes in a Simple IIB Orientifold'',
hep-th/0201028.},
\ref\frp{A.~R.~Frey and J.~Polchinski,
``${\cal N}=3$ Warped Compactifications'',
hep-th/0201029.}.\foot{These 
models are based on Type IIB orientifolds. 
For a very thorough introduction to orientifolds see
\ref\ASgnt{C.~Angelantonj and A.~Sagnotti,
``Open strings'', hep-th/0204089.} and
the references therein.} 
The potential
induced by the fluxes is of the no-scale type. This
implies that a non-vanishing (0,3)-flux breaks
supersymmetry \ref\GranaP{M.~Grana and J.~Polchinski, ``Supersymmetric
Three-Form Flux
Perturbations on AdS(5)'',
\physrev {63} {2001} {026001}, hep-th/0009211.}, \ref\Gubser{S.S.~Gubser,
``Supersymmetry and ${\cal F}$-Theory Realization of the
Deformed Conifold with Three-Form Flux'', hep-th/0010010.}
without introducing a vacuum energy.\foot{For the closely related
${\cal N} =2$ vacua
 the potential and low energy effective action were derived in
\ref\Michelson{
J.~Michelson,
``Compactifications of Type IIB Strings to Four Dimensions with Non-Trivial
Classical Potential'',
\np {495} {1997} {127}, hep-th/9610151.},
\ref\TV{T.R.~Taylor and C.~Vafa,
``R-R Flux on Calabi-Yau and Partial Supersymmetry Breaking'',
\pl {474} {2000} {130}, hep-th/9912152.},
\ref\Mayr{P.~Mayr,
``On Supersymmetry Breaking in String Theory and its Realization
in Brane Worlds'', \np {593} {2001} {99}, hep-th/0003198.},
\ref\ag{G.~Dall'Agata, ``Type IIB Supergravity Compactified on a
Calabi-Yau Manifold with H-Fluxes'', JHEP 0111:005 (2001), hep-th/0107264.},
\ref\lm{J.~Louis and A.~Micu, ``Type II Theories Compactified on Calabi-Yau
Threefolds in the Presence of Background Fluxes'', hep-th/0202168.}
and the vacuum structure was investigated in
\ref\CKLT{G.~Curio, A.~Klemm, D.~L\"ust and S.~Theisen,
``On the Vacuum Structure of Type II String Compactifications on
Calabi-Yau Spaces with H-Fluxes'',
\np {609} {2001} {3}, hep-th/0012213.}.}
Supersymmetry is generically broken at a scale that depends on the
volume of the Calabi-Yau threefold.
In
\ref\Beckero{K.~Becker and M.~Becker,
``Supersymmetry Breaking, ${\cal M}$-Theory and Fluxes'', JHEP 0107:038
(2001), hep-th/0107044.} an analogous version of this
mechanism was found for compactifications of
${\cal M}$-theory on eight-manifolds with
non-vanishing fluxes for tensor fields.\foot{The models found
in {\gkp} can be derived from {\Beckero} if the
eight-manifold is an elliptic fibration.}
The constraints imposed by supersymmetry on such compactifications
were determined in
\ref\Beckert{K.~Becker and M.~Becker,
``${\cal M}$-Theory on Eight-Manifolds'', \np {477} {1996} {155}
hep-th/9605053.}.
It was shown in
\ref\gvw{S.~Gukov, C.~Vafa and E.~Witten,
``CFT's from Calabi-Yau Four-Folds'', \np {484} {2000} {69},
hep-th/9906070.}
that these constraints can be derived from two
superpotentials that describe the vacuum solutions in three
dimensions.
These superpotentials and the corresponding scalar potential were
obtained by a Kaluza-Klein reduction of ${\cal M}$-theory
on a fourfold in
\ref\hlo{M.~Haack and J.~Louis, ``Duality in Heterotic Vacua with
Four Supercharges'', \np {575} {2000} {107}, hep-th/9912181.} and
\ref\hlt{M.~Haack and J.~Louis,
``${\cal M}$-Theory Compactified on Calabi-Yau Fourfolds with
Background Flux'',
\pl {507} {2001} {296}, hep-th/0103068.}.\foot{A similar Kaluza-Klein
reduction of Type IIA theory on Calabi-Yau fourfolds with background
fluxes is performed in \ref\HLM{M.~Haack, J.~Louis and M.~Marquart,
``Type IIA and Heterotic String Vacua in D = 2'', \np {598} {2001} {30}, hep-th/0011075.}, 
\ref\GH{
S.~Gukov and M.~Haack,
``IIA String Theory on Calabi-Yau Fourfolds with Background Fluxes'',
hep-th/0203267.} while the potential for ${\cal M}$-theory on
$G_2$-holonomy manifolds with fluxes has been computed in
\ref\bw{C.~Beasley and E.~Witten,
``A Note on Fluxes and Superpotentials in ${\cal M}$-theory
Compactifications
on Manifolds of $G_2$ Holonomy'', hep-th/0203061.}.
}

In this note we extend the analysis of {\gkp} and {\Beckero}
in two directions. First we rederive the results of {\Beckero} by
determining the minima of the potential calculated in {\hlt}.
This analysis will also show
that the non-supersymmetric vacua found in {\Beckero} are
classically stable. Our second aim is to investigate the fate of the
no-scale
structure of the potential if one considers the effect of higher derivative
corrections of string theory and ${\cal M}$-theory.
We confirm the expectation of {\gkp}
that the no-scale structure in Type IIB compactifications
 does not survive in the quantum theory  once higher
order $\alpha'$-corrections are taken into account.
In particular, this implies that
breaking supersymmetry via a (0,3)-flux induces a non-vanishing
potential. Due to the relationship between Type IIB
compactifications with three-form flux and ${\cal M}$-theory
compactifications
with four-form flux {\gvw}, \ref\DRS{K.~Dasgupta, G.~Rajesh and S.~Sethi,
``${\cal M}$-Theory, Orientifolds and G-Flux", JHEP 9908:023 (1999),
hep-th/9908088.} a similar result should be valid for the non-supersymmetric
fluxes
in ${\cal M}$-theory {\Beckero}, which also
lead to a vanishing cosmological constant at
leading order.

This paper is organized as follows. In section 2 we rederive the results of
{\Beckero} from the scalar potential derived in {\hlt} and show that the
non-supersymmetric vacua are classically stable. In section 3 we calculate
higher order $\alpha'^3$-corrections to the scalar potential computed in
{\gkp}. We show that these corrections generate a scalar
potential that depends on the Calabi-Yau volume and is
proportional to the Euler number of the internal manifold. This spoils the
no-scale structure of the classical scalar potential for manifolds with
non-vanishing Euler number and suggests that further $\alpha'$-corrections
may lead to a stabilization of the radial modulus.
Some of the technical details of the computation are relegated to
an appendix.

\newsec{(Non)-Supersymmetric Solutions in ${\cal M}$-theory}
In this section we derive the non-supersymmetric
vacuum solutions with vanishing cosmological constant
computed in {\Beckero} from the superpotentials found
in {\gvw} and {\hlt}.
We use the notation and conventions of
{\hlt}.

\subsec{The Scalar Potential}
The scalar potential of ${\cal M}$-theory
compactified on a fourfold $Y_4$ to three dimensions takes the
form\foot{We express the formulas in this section in terms of
the rescaled K\"ahler coordinates $\hat M^A = {\cal V}^{-1} M^A$
used in {\hlt}. However, in order not to overload the notation
we omit the hat on the coordinates here.}
\eqn\ai{V=e^{K^{(3)}}G^{-1\a{\bar \b}}D_{\a}WD_{\bar \b}{\bar W}
+({1\over 2} G^{-1AB}{\partial}_A{\hat W}{\partial}_B{\hat W}-
{\hat W}^2) \ ,}
where $\a,\b=1,\dots,  h^{3,1}$ label the complex structure
deformations while $A,B=1,\dots,h^{1,1}$ label the deformations of
the K\"ahler structure.
The K\"ahler potential\foot{Strictly speaking $K^{(3)}$ is not a
K\"ahler potential since in three dimensions the K\"ahler deformations
$M^A$ are not complexified. Nevertheless, the metric $G_{AB}$ is determined
by the second derivative of $K^{(3)}$.} $K^{(3)}$ is given in
terms of the holomorphic four-form $\Omega$ and the fourfold
volume ${\cal V}$
\eqn\aii{K^{(3)}=-\ln {\int}_{Y_4}\Omega \wedge { \bar \Omega}
+\ln {\cal V}\ ,}
while
\eqn\aiii{W={\int}_{Y_4} F \wedge \Omega}
and
\eqn\aiv{{\hat W}={1\over 4} {\int}_{Y_4}F\wedge J \wedge J}
represent the two superpotentials depending on the four-form flux $F$.
$G_{AB} = -{1 \over 2} \partial_A \partial_B K^{(3)}$ and
$G_{\alpha \bar \beta} = \partial_\alpha \partial_{\bar \beta} K^{(3)}$
are the
metrics on the moduli spaces of K\"ahler- and complex structure deformations
respectively and we further defined
$D_{\alpha}W={\partial}_{\alpha}W+({\partial}_{\alpha}K^{(3)})W$.

The scalar potential {\ai} originates entirely from the anti-selfdual
internal
components of the four-form $F$ {\hlt} as the selfdual part cancels out
from tadpole cancellation. To see this, expand the four-form flux as
$F = F_{4,0} + F_{3,1} + F_{2,2}
+ F_{1,3} + F_{0,4}$ and furthermore use the Lefschetz decomposition
in order to further expand $F_{2,2}$ as\foot{Note that this
decomposition slightly differs from the one in
{\hlt} in that we use the rescaled K\"ahler form here.}
\eqn\lefs{F_{2,2} =
F^{(0)}_{2,2} + J \wedge F^{(0)}_{1,1} + J^2 \wedge F^{(0)}_{0,0}\ .}
Here $F^{(0)}_{p,p}$ denotes a primitive $(p,p)$-form on $Y_4$, i.e.
it satisfies
\eqn\prim{J^{5-2p} \wedge F^{(0)}_{p,p} = 0\ .}
It has been shown in {\gvw} and {\hlt} that $F_{3,1}$,
$F_{1,3}$ and $J \wedge F^{(0)}_{1,1}$ are anti-selfdual, whereas the
other four-forms are selfdual. Furthermore, the potential
{\ai} can alternatively be rewritten as {\hlt}
\eqn\aitwo{V = - {\cal V} \Big( \int_{Y_4} F_{3,1} \wedge F_{1,3} + {1 \over
2}
\int_{Y_4} J \wedge F_{1,1}^{(0)} \wedge J \wedge F_{1,1}^{(0)} \Big)\ .}
Thus the potential only depends on the anti-selfdual part
of the four-form $F$.
It was noticed in {\gvw} that this scalar potential
is of no-scale type since it is independent of the volume
(up to an overall multiplicative factor coming from the Weyl-rescaling)
and  hence also does not depend on the radial
modulus of the fourfold.
The first term in the brackets of {\aitwo} is manifestly
volume independent. To see
this also for the second term, we rescale
$J$ by a factor $\lambda$. This amounts to rescaling the
volume by $\lambda^4$ and
requires also to rescale $F_{1,1}^{(0)}$  with $\lambda^{-1}$
(and $F_{0,0}^{(0)}$ with $\lambda^{-2}$)
in order to keep $F_{2,2}$ in {\lefs}
unchanged. Thus a rescaling of the volume precisely
drops out of the combination $F_{1,1}^{(0)} \wedge J$
and implies that also the second term in {\aitwo}
does not depend on the radial modulus.
Furthermore, $V$ is non-negative which can be seen by rewriting {\ai}
once more as
\eqn\aithree{V=e^{K^{(3)}}G^{-1\a{\bar \b}}D_{\a}WD_{\bar \b}{\bar W}
+{1\over 2} G^{-1AB}D_A{\hat W}D_B{\hat W}\ ,}
where we introduced the covariant derivative
$D_A \hat W = \partial_A  \hat W - {1 \over 2} (\partial_A K^{(3)}) \hat W$.
Written in this form  $V$ is manifestly non-negative.

\subsec{Supersymmetry}
In supergravity theories with four supercharges the conditions
for unbroken supersymmetry for compactifications to three-dimensional
Minkowski space are {\gvw}
\eqn\av{W=D_{\alpha}W=0\ ,}
and
\eqn\avi{{\hat W}={\partial}_A{\hat W}=0\ .}
The constraints for a supersymmetric three-dimensional vacuum
found in {\Beckert} then easily follow from the above conditions
{\gvw}, {\hlt}. First, $W=0$ implies $F_{4,0}=F_{0,4}=0$. Second,
the identity
$D_{\alpha}W=\int {\Phi}_{\a} \wedge F $, where
${\Phi}_{\a}$ is a basis of $(3,1)$-forms, gives together with
$D_{\alpha}W=0$ the condition $F_{1,3}=F_{3,1}=0$.
Third, the condition ${\partial}_A{\hat W}=0$ implies that $F$
is primitive
\eqn\avii{F\wedge J=0\ ,}
while ${\hat W}=0$ follows from the above condition and imposes
no new constraint.

Let us now consider the non-supersymmetric solutions.
Taking the leading quantum gravity corrections of ${\cal M}$-theory
into account it was shown in {\Beckero} that turning on a flux of the
form
\eqn\aviii{F=F_{4,0}+F_{0,4}+F_{0,0}J \wedge J\ ,}
breaks supersymmetry without generating a cosmological
constant. This can also easily be seen from the previous discussion.
From the potential \aithree\ we see that the solution to the 
equations of motion has to satisfy
\eqn\newsol{
D_{\alpha} W= D_A {\hat W}=0.
}
The vanishing of $D_{\alpha}W$ implies $F_{1,3}=F_{3,1}=0$,  
while $D_A {\hat W}=0$ implies $F_{1,1}^{(0)}\wedge J=0$. Therefore 
we see that the flux has to be selfdual. It is clear from 
\aithree\ that theses solutions do not generate a cosmological 
constant. A flux of the form 
$F=F_{4,0}+F_{0,4}$ breaks supersymmetry as $W\neq 0$, while 
a flux $F\sim J \wedge J$ is not primitive implying 
${\hat W} \neq 0$ and supersymmetry is broken in this case as 
well. 
These non-supersymmetric vacua are stable to this order as the potential
{\ai}
is non-negative. The solution $V=0$ then corresponds to a
minimum of the potential and a vanishing of the cosmological constant.
To summarize,
the non-supersymmetric compactifications
of ${\cal M}$-theory with a vanishing three-dimensional cosmological
constant found in {\Beckero} are stable to leading
order as they minimize
the supergravity potential.
The equation {\newsol} is the determining
equation for the moduli fields. Generically it should be possible to
stabilize all of the moduli fields appearing in
these compactifications but for the radial modulus as the
superpotential is independent of this modulus.
This is in contrast to the Type IIB theory
considered in {\gkp}, where only a superpotential for the complex
structure moduli is generated and none of the K\"ahler moduli fields
can be stabilized.

Once supersymmetry is broken by the flux {\aviii} the gravitino
becomes massive. To see this we notice that
the relevant terms
of the action of eleven-dimensional supergravity
\ref\cjs{E.~J.~Cremmer, B.~Julia and L.~Scherk, ``Supergravity Theory in
Eleven Dimensions'',
 \pl {5} {1978} {409}.} are
\eqn\bi{L=-{{\sqrt 2}{\kappa}\over 384}e(\bar {\Psi}_M
{\G}^{MNPQRS}{\Psi}_S+12\bar {\Psi}^N{\G}^{PQ}
{\Psi}^R)F_{NPQR}\ .}
In order to compactify this interaction to three dimensions
we decompose the eleven-dimensional gamma matrices as in
{\Beckert} while the eleven-dimensional gravitino is
decomposed as
\eqn\bii{{\Psi}_{\m}={\psi}_{\m}\otimes {\epsilon}+
{\psi^*}_{\m}\otimes {\epsilon}^*+\dots ,}
where ${\psi}_{\m}$ is the three-dimensional gravitino and
${\epsilon}$ is a complex eight-dimensional
spinor on the internal manifold with unit norm. In the above formula
there are further terms that are necessary in order to eliminate the
mixed components of the gravitino mass matrix.
Taking this decomposition into account
we obtain the following mass terms for the gravitino\foot{Strictly 
speaking the correct mass terms are obtained after the Weyl rescaling to 
the Einstein frame.}
\eqn\biii{{\bar {\psi^*}_{\m}}{\Gamma}^{\m\n}{\psi}_{\n}{\int}_{Y_4}
{\Omega}\wedge F+c.c. \,}
for $F$ of type $(0,4)$, while for $F$ non-primitive
the mass term becomes
\eqn\biv{{\bar {\psi}_{\m}}{\Gamma}^{\m\n}{\psi}_{\n}{\int}_{Y_4}
J \wedge J \wedge F +c.c. \ .}
Here we have defined ${\bar \psi}_{\m}(x)={\psi}_{\m}^T{\G}_0$.
{}From the interactions {\biii} and {\biv} we see that the gravitino
mass vanishes for supersymmetric compactifications in which $F_{0,4}=F_{4,0}=0$ 
and $F\wedge J=0$ holds respectively.

\newsec{Higher Order Corrections to the Potential in Type IIB Theory}

Our next goal is to analyze the effects of higher order terms
(like $F^2R^3$ terms) appearing in the ${\cal M}$-theory
effective action on the scalar potential. Unfortunately
only a few results are known about these terms so that
a direct derivation of the corrections to the potential via a
Kaluza-Klein reduction is unfeasible at present.
The situation for the Type IIB theory is not much better.
Some of the relevant terms have been computed in
\ref\kepa{A.~Kehagias and H.~Partouche, ``The Exact Quartic Effective Action
for the Type IIB Superstring'', \pl {422} {1998} {109}, hep-th/9710023.},
\ref\pvwo{K.~Peeters, P.~Vanhove and A.~Westerberg, ``Supersymmetric Higher
Derivative Actions
in Ten Dimensions and Eleven Dimensions, the Associated
Superalgebras and their Formulation in Superspace'', \cqg {18} {2001}
{843}, hep-th/0010167.},
\ref\fkt{S.~Frolov, I.~Klebanov and A.~Tseytlin, ``String Corrections
to the Holographic RG Flow of Supersymmetric
$SU(N)\times SU(N+M)$ Gauge Theory'', \np {620} {2002}
{84}, hep-th/0108106.} and
\ref\pvwt{K.~Peeters, P.~Vanhove and A.~Westerberg,
``Chiral Splitting and
World Sheet Gravitinos
in Higher Derivative String Amplitudes'', hep-th/0112157.}
but a complete calculation of all the leading higher order interactions
is still lacking.\foot{For a review with a complete list of references see
\ref\green{M.B.~Green, ``Interconnections Between Type II Superstrings,
${\cal M}$-theory and $N=4$ Supersymmetric Yang-Mills'', hep-th/9903124.}.
The
effect of such higher order interactions on the dual
confining gauge theory has recently been analyzed in
{\fkt}.}
As mentioned in the introduction, in the Type IIB theory compactified
on a Calabi-Yau threefold with fluxes (and external brane sources)
a potential of no-scale type is also induced {\gkp} and its structure is
similar to the potential discussed in the previous section.
Correspondingly  it was conjectured that ${\alpha}'$- and string
loop-corrections generate a potential for the radial modulus so that
the no-scale structure is lost {\gkp}.
Fortunately, in these compactifications there is a different way to
obtain the ${\alpha}'$-corrections to the potential
which uses mirror symmetry and the c-map of \ref\fs{S.~Ferrara
and S.~Sabharwal, ``Quaternionic Manifolds for Type II Superstring
Vacua of Calabi-Yau Spaces'', \np {332} {1990} {317}.}.

Let us first recall
that the metric for the K\"ahler deformations of the
threefold
receives perturbative ${\alpha}'^3$-corrections from
higher derivative terms appearing in the ten-dimensional Type II effective
action
(at tree-level the relevant terms coincide for the Type IIA and IIB theory).
This has been shown in
\ref\afmn{I.~Antoniadis, S.~Ferrara, R.~Minasian and
K.~S.~Narain, ``$R^4$ Couplings in ${\cal M}$ and Type II
Theories on Calabi-Yau
Spaces'', \np {507} {1997} {571}, hep-th/9707013.}
using the results of \ref\gw{D.~Gross and E.~Witten,
``Superstring Modification of
Einstein Equations'',  \np {277} {1986} {1}.},
\ref\fp{M.D.~Freeman and C.N.~Pope,
``Beta-Functions and Superstring Compactifications'',
 \pl {174} {1986} {48}.},
\ref\grisarut{M.~T.~Grisaru, A.~E.~M.~van de Ven and D.~Zanon, ``Four-Loop
Divergences for the $N=1$ Supersymmetric Non-Linear Sigma-Model
in Two Dimensions'', \np {277} {1986} {409}.} and
\ref\cogp{P.~Candelas, X.~C.~De La Ossa,
P.~S.~Green and L.~Parkes, ``A Pair of Calabi-Yau Manifolds as an
Exactly Soluble Superconformal Field Theory'', \np {359} {1991} {21}.}.
The relevant terms in the Type II effective action
responsible for the correction of the K\"ahler moduli space metric are
\eqn\corr{S=-{1\over {2{\kappa}^2_{10}}}
\int d^{10} x \sqrt{-g^{(10)}}
e^{-2
\phi} (R+4(\partial \phi)^2 +\alpha'^3 c_1 J_0 )\ ,}
where $c_1={\zeta(3)\over 3\cdot 2^{11}}$. The higher order
interaction is defined as
\eqn\jzero{ J_0 = t^{M_1 N_1 \ldots M_4 N_4} t_{M_1' N_1' \ldots
M_4' N_4'} R^{M_1' N_1'}\! _{M_1 N_1} \ldots R^{M_4' N_4'}\! _{M_4
N_4} + {1 \over 4} E_8\ , }
where capital letters indicate ten-dimensional indices and $\phi$
is the ten-dimensional Type II dilaton.
The tensor $t$ is defined as in {\fkt} and we are using these
conventions in the following.
$E_8$ is a ten-dimensional generalization of the eight-dimensional
Euler density given by
\eqn\eeight{E_8 = {1 \over 2} \epsilon^{ABM_1 N_1 \ldots M_4 N_4}
\epsilon_{ABM_1' N_1' \ldots M_4' N_4'} R^{M_1' N_1'}\! _{M_1 N_1}
\ldots R^{M_4' N_4'}\! _{M_4 N_4}\ . }
In the appendix we show that we also need a term
\eqn\ddil{\sim {1\over {2{\kappa}^2_{10}}}
\int d^{10} x \sqrt{-g^{(10)}} e^{-2 \phi} \alpha'^3 (\nabla^2 \phi) Q,}
where $Q$ is a generalization of
the six-dimensional Euler integrand, $\int_{Y_3} d^6x \sqrt{g} Q = \chi$,
which is explicitly defined in (A.5).
This term does not
modify the equations of motion
to order ${\cal O}(\alpha'^3)$ but is necessary in order to derive the
correct four-dimensional low energy
effective action to that order. This is shown in detail in the appendix.

After compactification on a Calabi-Yau threefold the interactions {\corr}
together with {\ddil} give the perturbative correction to
the metric on the moduli space of the K\"ahler deformations
computed in {\cogp}. In particular, it has been shown in
{\cogp} how this correction modifies the prepotential of the
K\"ahler moduli space in Calabi-Yau compactifications of the
Type II theories. In the Type IIB case the K\"ahler deformations reside
in ${\cal N}=2$ hypermultiplets. However, we are actually interested in
orientifold and ${\cal F}$-theory compactifications, so that part of the
fields appearing in these compactifications have to be projected
out.\foot{Strictly speaking our formulas only apply to the
Calabi-Yau orientifold case discussed in {\gkp}.}
More precisely in the Type IIB hypermultiplet moduli space
we need to perform a truncation to an ${\cal N}=1$ subsector
by projecting out the moduli that arise from the ten-dimensional
antisymmetric two-forms. 

We first need to know the corrections
to the metric on the hypermultiplet moduli space. It has been shown in
{\fs} that this metric is entirely expressible in terms of the
prepotential, via the c-map. Thus all the perturbative corrections
to the hypermultiplet moduli space are captured by those to the
prepotential calculated in {\cogp}. However, the hypermultiplet moduli space
has been parametrized in {\fs} in variables whose relation to the
ten-dimensional Type IIB fields, on which the truncation acts naturally,
are not obvious. In fact the transformation
relating the two field bases is rather involved and has been established
in \ref\bghl{R.~B\"ohm, H.~G\"unther,
C.~Herrmann and J.~Louis, ``Compactification of Type IIB String
Theory on Calabi-Yau Threefolds'', \np {569} {2000} {229},
hep-th/9908007.}.\foot{For the special case $h^{1,1}=1$ see also
\ref\bc{M.~Bodner and A.~C.~Cadavid, ``Dimensional Reduction of the
Type IIB Supergravity and Exceptional Quaternionic Manifolds'', \cqg {7}
{1990} {829}.}.} Thus we first have to translate the
perturbative corrections to the hypermultiplet action
into the appropriate Type IIB variables.

Let us start from the action of the hypermultiplets in
Type IIB compactifications on the
Calabi-Yau threefold $Y_3$ given in {\fs}\foot{We have adjusted the
formula of {\fs} to our conventions, i.e.\ the +++ conventions
in the language of \ref\mtw{C.W.~Misner, K.S.~Thorne and J.A.~Wheeler, ``Gravitation'',
{\it Freeman, San Francisco} (1973).}.}
\eqn\cmap{\eqalign{- \sqrt{-g}^{-1} L = & {1 \over 2} R - G_{a \bar
b}(z, \bar z) \partial^\mu z^a \partial_\mu \bar z^{\bar b} -
(\partial^\mu \phi_B)^2 \cr & - {1 \over 4} e^{4 \phi_B}
\Big(\partial_\mu \tilde \phi + \zeta^i
\partial_\mu \tilde \zeta_i - \tilde \zeta_i
\partial_\mu \zeta^i \Big)^2 +{1 \over 2} e^{2 \phi_B}
\partial_\mu \zeta^i
R_{ij}(z, \bar z) \partial^\mu \zeta^j \cr & +{1 \over 2} e^{2
\phi_B} \Big(I_{ik}(z, \bar z)
\partial^\mu \zeta^k + \partial^\mu \tilde \zeta_i\Big)
R^{-1ij}(z, \bar z) \Big(I_{jl}(z, \bar z)
\partial_\mu \zeta^l + \partial_\mu \tilde \zeta_j\Big)\ ,}}
where the $z^a , \ a = 1, \ldots ,h^{1,1}(Y_3),$
are the K\"ahler deformations of $Y_3$.
They are related to the projective coordinates
$X^i, \ i = 0, \ldots ,h^{1,1}(Y_3)$, according to
\eqn\prcoord{z^i= {X^i \over X^0}\, \quad ({\rm i.e.}\, \ z^0=1)\ .}
The $\zeta^i$, $\tilde \zeta^i$ and $\tilde \phi$ in {\cmap}
are related to the scalars arising in the R-R sector and the scalar dual to
the NS-NS antisymmetric tensor in a complicated way {\bghl}, whereas
$\phi_B$ is the four-dimensional Type IIB dilaton.
All couplings are determined in terms of a holomorphic
prepotential $F(X)$ via
\eqn\notation{\eqalign{R_{ij} & = {\rm Re}\,\CN_{ij}\ ,\qquad I_{ij} = {\rm
Im}\,\CN_{ij}\ ,
\cr \CN_{ij} & = {1 \over 4}  \bar{F}_{ij} - {(Nz)_i (Nz)_j \over (zNz)}\ ,
\cr F_{ij} & = {\partial^2 F \over \partial X^i \partial X^j}\ ,
\qquad N_{ij} =  {1 \over 4} ( F_{ij} + \bar{F}_{ij} )\ ,
\cr (Nz)_i & = N_{ij} z^j\ , \qquad (zNz) = z^i N_{ij} z^j\ ,}}
where we use the conventions and notation of {\bghl}.
The metric $G_{a \bar b}$ is K\"ahler  with a K\"ahler potential
\eqn\k{K = -\ln[X^i \bar F_i(\bar X) + \bar X^i F_i(X)]\ ,}
that is also expressed in terms of $F$.
The components of the metric then take the form
\eqn\gab{G_{a \bar b} = {\partial^2 K \over
\partial z^a \partial \bar z^b} = - {1 \over z N \bar z} \left( N_{ab}
- {(N \bar z)_a (N z)_b \over (z N \bar z)} \right)\ .}

The prepotential for the K\"ahler deformations
receives perturbative and non-per\-tur\-ba\-tive corrections
on the worldsheet. These have been successfully computed {\cogp}
using mirror symmetry and the perturbative correction
has been identified with the $\alpha'$-corrections
determined in {\gw}, {\fp},
\ref\grthree{M.T.~Grisaru, A.E.~van de Ven and D.~Zanon,
``Four Loop Beta Function for fhe ${\cal N}=1$ and ${\cal N}=2$
Supersymmetric Nonlinear Sigma Model in Two Dimensions'', \pl {173} {1986}
{423}}
and \ref\grtwo{M. T. Grisaru, A.E.M.
van de Ven and D. Zanon,
``Two-dimensional Supersymmetric Sigma Models on Ricci Flat K\" ahler
Manifolds are not Finite'', \np {277} {1986} {388}.}. The perturbative
prepotential
determined in this way reads
\eqn\prepot{F(X) = {i \over 3!} \kappa_{abc} {X^a X^b X^c \over X^0} +
(X^0)^2 \xi\ .}
The cubic term is the classical contribution with
 ${\kappa}_{abc}$ being the classical
intersection numbers of $Y_3$. The constant term is
proportional to the Euler number $\chi$ of $Y_3$ and in the
appendix we derive its precise value
\eqn\xiv{\xi = -{\chi\over 2} \zeta(3)\ .}
It describes the (worldsheet) perturbative quantum corrections.
Inserting \prepot\ into \k\ leads
to the corrected K\"ahler potential of the K\"ahler class deformations
$z^a$
\eqn\kcorr{K =  -\ln[-{i \over 6} \kappa_{abc} (z^a - \bar z^a)
(z^b - \bar z^b) (z^c - \bar z^c) + 4 \xi]\ . }

Finally we need to display
the hypermultiplet action {\cmap}
in terms of the Type IIB field variables
using the explicit map given in {\bghl}.\foot{The map in {\bghl}
has been interpreted there
as the mirror map relating the Type IIA and Type IIB
hypermultiplet sectors. It can,
however, equally well been understood just inside the Type IIB
theory, relating two coordinate
bases of the hypermultiplet sector, the one used in {\fs} and
the one which naturally arises
in a Kaluza-Klein reduction.}
However, here  we are interested in truncating the
spectrum to an ${\cal N}=1$ subsector in which the fields
coming from
the Type IIB antisymmetric two-forms are projected out.
In this subsector the map from \bghl\ takes
the form
\eqn\map{\eqalign{\zeta^0 & = \sqrt{2} l\ , \cr
\tilde \zeta_a & = - {\sqrt{2}  \over 4 } g_a\ , \cr
\phi_B & = \phi_4\ , \cr
{\rm Im}(z^a) & = -v^a\ ,}}
with all other fields being projected out. In this formula
we have used the notation of
{\bghl} in which $l$ is the R-R scalar of Type IIB,
$\phi_4$ the four-dimensional Type IIB
dilaton, $g_a$ are the scalars dual to the antisymmetric tensors
coming from
expanding the four-form into a basis of (1,1)-forms and $v^a$ are the
K\"ahler class moduli of $Y_3$.
These fields describe a set of $2(h^{1,1}+1)$ real coordinates.
Since the projection {\map} breaks ${\cal N}=2$ supersymmetry to
${\cal N}=1$,
the quaternionic geometry of the hypermultiplet moduli space
must reduce to a K\"ahler geometry.\foot{The general situation of
truncating an ${\cal N} =2$ theory to ${\cal N} =1$
is discussed in \ref\adf{
L.~Andrianopoli, R.~D'Auria and S.~Ferrara,
``Supersymmetry Reduction of ${\cal N}$-Extended Supergravities
in Four Dimensions'',
JHEP 0203:025 (2002), hep-th/0110277.} and
\ref\AndrianopoliGM{
L.~Andrianopoli, R.~D'Auria and S.~Ferrara,
``Consistent Reduction of ${\cal N} = 2 \to {\cal N} = 1$
Four Dimensional Supergravity
Coupled to Matter'', \np {628} {2002} {387}, hep-th/0112192.}. We
have checked that the projection {\map} is consistent with the
formalism described in {\adf}.}
Hence our next task is to display appropriate
complex coordinates in which the truncated hypermultiplet
metric is manifestly K\"ahler.
To lowest order in $\alpha'$ and at the string tree-level this
has  already been done  in {\gkp}.
The relevant field variables for the case of only one K\"ahler modulus
were found to be
$\tau = l + i e^{-\phi_0}$ with $\phi_0$ being
the ten-dimensional Type IIB dilaton to leading order
and $\rho = {1\over 3}g + i e^{4u}$ with $e^{6u}$ being
the volume of $Y_3$ in the Einstein frame.
Our goal now is to find a definition of $\tau$ and
${\rho}$ which takes into account the higher order correction
appearing in the prepotential {\prepot}
and which
is valid for the case of more than one K\"ahler modulus.
In order to do so we need to express the four-dimensional
dilaton appearing in {\map} in terms of the ten-dimensional
dilaton.
{}From the work {\gw},
\ref\sen{A.~Sen,
``Central Charge of the Virasoro Algebra for
Supersymmetric Sigma Models on Calabi-Yau Manifolds'', \pl {178} {1986}
{370}.},
\ref\fpss{
M.D.~Freeman, C.N.~Pope, M.F.~Sohnius and
K.S.~Stelle, ``Higher Order $\sigma$-Model
Counterterms and the Effective Action
for Superstrings'', \pl {178} {1986} {199}.},
\ref\cfpss{P.~Candelas, M.D.~Freeman, C.N.~Pope, M.F.~Sohnius and
K.S.~Stelle, ``Higher Order Corrections to Supersymmetry and
Compactifications of
the Heterotic String'', \pl {177} {1986} {341}.}
and {\afmn}
it is known that the equation
of motion for the ten-dimensional dilaton gets modified in the presence of the
higher derivative term {\corr}
such that a constant dilaton is not a solution anymore.
Rather,
the solution
becomes
\eqn\dilaton{\phi = \phi_0 +{ \zeta(3) \over 16} \ Q\ ,}
where $\phi_0=\phi_0(x)$ is the uncorrected, constant dilaton and
$Q$ is defined in (A.5). The value of the constant appearing
in the correction term of {\dilaton} is determined in the appendix.
Here and in the following we set
$2\pi {\alpha}'=1$ which implies
\ref\joep{J.~Polchinski, ``String Theory'', 2 vols.,
{\it Cambridge University Press} (1998).}
\eqn\nc{2{\kappa_{10}}^2=(2\pi)^7(\alpha')^4=(2\pi)^3.}
The ${\alpha}'$-dependence of our formulas can be easily restored at the end
by dimensional analysis.
In order to find the
right definition of the four-dimensional dilaton
in terms of the ten-dimensional dilaton to order
${\cal O}(\alpha'^3)$ one has to
compactify the action {\corr} together with {\ddil} in the background
\dilaton\ and determine the function in front of the curvature scalar in
four dimensions.
This is done in the appendix and leads to the following
definition of the four-dimensional dilaton
to order ${\cal O}(\alpha'^3)$
\eqn\fourdil{e^{-2 \phi_4} = e^{-2 \phi_0} \Big({\cal V} + {1\over2} \xi
\Big)}
in terms of the Calabi-Yau volume ${\cal V}$ and the higher
order correction
$\xi$.
Here and in the following we are using the notation
\eqn\volume{\eqalign{
& {\cal V}={1 \over 6} \kappa_{abc} v^a v^b v^c , \cr
& {\cal V}_a= {1 \over 6} \kappa_{abc} v^b v^c , \cr
& {\cal V}_{ab}= {1 \over 6} \kappa_{abc}  v^c .\cr}
}
Transforming {\fourdil} into the Einstein frame gives
\eqn\fdilein{e^{-2 \phi_4} = e^{-1/2 \phi_0} \Big(\CV
+{1 \over 2} \xih  \Big)\ ,}
where we have defined
${\hat \xi}=\xi e^{-3/2 \phi_0}$ and used
\eqn\ekahler{v^a = \hv^a e^{\phi_0/2}}
in order to relate the K\"ahler moduli in the string frame to those
in the Einstein
frame.\foot{Our definition of the ten-dimensional
Einstein frame proceeds
via a Weyl-rescaling $g_{MN}^{(S)} = e^{1/2 \phi_0}
g_{MN}^{(E)}$.}
These are the relevant variables in order to make
contact with {\gkp}.
Without the higher derivative correction ${\rm Im} (\tau)$
was defined as the four-dimensional fluctuations around the
constant dilaton
background $\phi_0$. We keep this definition of $\tau$ in
the following
and continue to define $\tau = l + i e^{-\phi_0}$ as in {\gkp}.

Inserting \map , \fdilein\ and \ekahler\ into \cmap\ we arrive
at the following Lagrangian for
the ${\cal N}=1$ truncation
\eqn\trunc{\eqalign{
- \sqrt{-g}^{-1} L = &  {1 \over 2} R +
{2e^{\phi_0/2}\o \xih +2 \vh} \left[  R_{00}
((\partial_{\mu} l)^2+e^{-2 \phi_0 }(\partial_{\mu} \phi_0 )^2 )+
R^{-1}_{ab}(x_a x_b+y_a y_b )\right], \cr }
}
where
\eqn\aai{\eqalign{
& x_a=I_{a0}\partial_\mu l-{1\over 4} \partial_{\mu }g_a\ , \cr
& y_a=-e^{-\phi_0}I_{a0}\partial_{\mu }\phi_0 -{3\over 4}\partial _{\mu}
\vh_a. \cr}}
It is straightforward to compute the non-vanishing
components of the couplings
\eqn\ir{
\eqalign{
I_{a0} & ={9  \over 4}e^{\phi_0}\left( {\xih \vh_a  \over
\xih -4\vh }\right) \ , \cr
R_{00} & ={1\over 2} e^{3 \phi_0/2}\left({2 \vh^2-\vh
\xih -\xih^2\over
\xih -4 \vh}\right)
\ ,
\cr
R_{ab} & = {3\over 2} e^{\phi_0/2} \left( \vh_{ab}+6
{\vh_a \vh_b \over
\xih-4 \vh }\right)
\ ,}}
while all other components vanish.
The sigma model metric displayed in {\trunc}
is indeed K\"ahler which becomes manifest in
the K\"ahler coordinates\foot{The indices are raised with
$\delta^{ab}$.}
\eqn\kcoord{\eqalign{T^a & = {1\over 3} g^a + i \CV^a\ , \cr
\tau & = l + i e^{-\phi_0}\ .}}
The corrected K\"ahler potential takes the following form
\eqn\kpot{\eqalign{{\cal K} & = -\ln [-i(\tau - \bar \tau)]
- 2 \ln\left(-i (T^a - \bar T^a) \hat v_a
+ \xi \Big({-i(\tau - \bar \tau) \over 2}\Big)^{3/2}\right)
-\ln[-i \int_{Y_3} \Omega \wedge \bar \Omega]
\cr & = \phi_0 - 2 \ln(\CV + {1 \over 2} \xi e^{-3 \phi_0 /2})
-\ln[-i \int_{Y_3} \Omega \wedge \bar \Omega] + {\rm const.}\ ,}}
where $\hat v_a$ in the first line is understood to be a function
of $-i(T^a - \bar T^a)$ given by the inverse of \volume .
The last term in the above expression is
the K\"ahler potential for the complex structure moduli.\foot{In the case
of Calabi-Yau orientifolds the complex structure moduli are restricted
to those even under the orientifolding {\gkp}.}

Before we proceed to determine the corrections to the
supergravity potential
implied by the corrections to the K\"ahler potential,
let us discuss
the symmetries of the four-dimensional effective theory.
These are important for arguing that all $\alpha'$-corrections
to the potential
originate from a correction to the K\"ahler potential, while the
superpotential is not corrected.
The real parts of the K\"ahler coordinates $T^a$ originate from
the ten-dimensional R-R four-form and inherit from its gauge
invariance
a Peccei-Quinn (PQ) shift symmetry, which
is not broken by the $\alpha'^3$-corrections {\gkp}. Also $\tau$
has a shift symmetry
which is a special case of the $SL(2, \ZZ)$ symmetry of the
Type IIB theory. However, the $SL(2, \ZZ)$
symmetry naturally
acts on the ten-dimensional dilaton $\tau_{10} = l + i
e^{-\phi}$ according to
\eqn\sltz{\tau_{10} \rightarrow \tau'_{10}={a \tau_{10} +
 b \over c \tau_{10} +d}\ ,
\qquad a, \ldots, d \in  \ZZ\ , \qquad ad -bc =1\ , }
and therefore has no obvious action on the four-dimensional
 $\tau$ defined in {\kcoord}.
The origin of this problem is that the derivation of {\trunc}
and therefore of the K\"ahler coordinates {\kcoord} and the
K\"ahler potential {\kpot} implicitly makes use of a
background for the
ten-dimensional dilaton given by {\dilaton} and a
constant $l$. However,
this configuration is not invariant
under a general $SL(2, \ZZ)$ transformation as only the subgroup
of {\sltz} with $c=0$ leaves $l$ constant. This subgroup is still
manifest in the four-dimensional effective theory (including the
$\alpha'^3$-corrections
to the potential) and in particular contains the shift symmetry of $l$.

We now compute the form of the corrected supergravity potential.
The ${\cal N}=1$ scalar potential reads
\eqn\potfour{V = {e^{\cal K}\over 2{\kappa^2_{10}}}
 \Big( G^{-1 I \bar J} D_I W
D_{\bar J} \bar W - 3 |W|^2 \Big)\ ,}
where $I, J$ label the
scalar fields of the theory.
The K\"ahler covariant derivatives are given by
$D_I W = (\partial_I + (\partial_I{\cal  K})) W$.
In {\gkp} it was shown that to leading order, i.e.\ for $\xi=0$,
and $h_{11}=1$ 
the potential derived via a Kaluza-Klein reduction is
indeed of the form {\potfour} with the K\"ahler potential
{\kpot} evaluated at $\xi=0$. The superpotential for the complex
structure moduli fields is
\eqn\super{W = \int_{Y_3} G_{(3)} \wedge \Omega\ .}
This superpotential has first been derived in {\TV} in the context of ${\cal N} = 2$
compactifications of the Type IIB theory with fluxes.
Here $\Omega$ is the $(3,0)$-form of the Calabi-Yau manifold $Y_3$
and $G_{(3)}$ is the complex three-form
\eqn\gthree{G_{(3)} = F_{(3)} - \tau H_{(3)}\ ,}
which transforms as
\eqn\gtrafo{G_{(3)} \rightarrow {G_{(3)} \over c \tau + d}}
under an $SL(2, \ZZ)$-transformation {\sltz}.\foot{Note that
to lowest order in $\alpha'$ the dilaton $\tau$ coincides with
$\tau_{10}$.} At the same time the K\"ahler potential
{\kpot} at $\xi=0$ transforms as
\eqn\ksltz{{\cal K}^{(0)} \rightarrow {\cal K}^{(0)}
+ \ln(c \tau + d) + \ln(c \bar \tau + d)\ .}
Thus to lowest order in $\alpha'$
an $SL(2, \ZZ)$-transformation {\sltz} acts as a K\"ahler transformation
in the low energy effective action, which is therefore left invariant.
Thus the fact that the shift symmetry of $\tau$ is actually a special case
of
an $SL(2, \ZZ)$-transformation is the reason why the dilaton is allowed
to appear in the superpotential {\super}, albeit only in the combination
{\gthree}.

Generalizing to $\xi\neq 0$ we note that
{\super} is still the relevant superpotential in this case,
such that
all the corrections to the potential come from corrections to the
K\"ahler potential {\kpot}. This is because
the superpotential does not receive any $\a'$-corrections,
as has been argued in {\gkp} using the argument of
\ref\newissues{
E.~Witten, ``New Issues in Manifolds of $SU(3)$ Holonomy'',
\np {268} {1986} {79}.} that the PQ symmetry
of the K\"ahler moduli $T^a$ forbids their appearance in the
superpotential. However,
as in the lowest order case the appearance of the dilaton in the
combination {\gthree} is possible because its unbroken shift-symmetry
is a special case of an $SL(2, \ZZ)$-transformation.
Thus {\super} should still be the relevant superpotential in
the case at hand.

It was shown in {\gkp} that to leading order in $\a'$ the
$W^2$-term in {\potfour} cancels out leaving a
non-negative potential of no-scale type. At the minimum of
this potential (i.e.\ at $V=0$)
the Type IIB complex structure moduli can be fixed but all
the K\"ahler moduli
(including the radial modulus) remain undetermined {\gkp}.
In order to compute the corrections to the potential 
we can use the superspotential {\super} in {\potfour}
and the corrected K\"ahler potential {\kpot}.
%
%
This leads to the following form of the potential
\eqn\potential{
\eqalign{ 
V & = {e^{\cal K}\over 2{\kappa^2_{10}}} \Big[(G^{-1})^{\alpha 
{\bar \beta}}D_{\alpha} W 
D_{\bar \beta} {\bar W} +(G^{-1})^{\tau {\bar \tau}}D_{\tau} W 
D_{\bar \tau} {\bar W} 
\cr
& -9 { \xih \vh e^{-\phi_0}\over (\xih-\vh)(\xih+2 \vh)}(W D_{\bar \tau} 
{\bar W} +{\bar W}D_{\tau}W)
 -3 \xih {\xih^2+7 \xih \vh+\vh^2\over (\xih-\vh)(\xih+2\vh)^2}
|W|^2\Big] .
}}
Here we have restored the dependence on Newton's constant.
Clearly the modified potential does not exhibit the no-scale structure of
the
tree-level potential any more due to its non-trivial dependence on the
radial
mode $\rho$ and in particular the $|W|^2$-term does not drop out.
This means that breaking supersymmetry via\foot{Note that the supersymmetry
conditions arising from $D_I W=0$ are the same as to lowest order. In
particular
they still demand $W=0$.} $W \neq 0$ does not lead to
a vanishing vacuum energy any more.
We expect that a similar result is valid for the
non-supersymmetric fluxes
in ${\cal M}$-theory {\Beckero}, which also
lead to a vanishing cosmological constant at
leading order. This is suggested by the relationship between Type IIB
compactifications with three-form flux and ${\cal M}$-theory
compactifications
with four-form flux {\gvw}, {\DRS}.

Further corrections to the K\"ahler potential
may arise due to the presence
of the orientifold O3-planes and D3-branes. Certainly D3-branes
would give rise to
additional moduli that we did not take into account. It is
however possible
to cancel the O3-plane charge totally by the flux, so that
in this case
no D3-branes have to be included {\KST}. The only effects
of the localized
sources that we make use of are that they break supersymmetry by
one half and
that they modify the Bianchi identity for the 
five-form field strength. In this way
they guarantee the possibility to turn on non-trivial fluxes. 
However, introducing the localized sources also
leads to a non-trivial warp factor. It has been argued in {\gkp} and {\frp} that
its effect is subleading in the large volume limit. But also the $\alpha'^3$
corrections to the potential are suppressed in this limit, c.f.\ (3.32).
It would be interesting to get a better understanding of the effects of 
a non-trivial warp factor on the potential.

As the Type IIB theory also contains correction terms of higher order in
$\alpha'$ than
${\cal O}(\alpha'^3)$ we can probably trust our result {\potential} only to
order
${\cal O}(\alpha'^3)$. Let us therefore explicitly exhibit the correction
terms up
to that order compared to the tree-level potential. Expanding {\potential}
they are found to be
\eqn\coralthree{
\delta V =
-{{\hat \xi} \over \hat {\cal V}} V_{\rm tree}+
{3\o 8} { e^{{\cal K}^{(0)}}\over \kappa_{10}^2}
{{\hat \xi} \over {\hat {\cal V}}}
\mid W+(\tau -{\bar \tau}) {\tilde D}_{\tau}W\mid^2\ ,
}
where ${\cal K}^{(0)}$ is the tree-level K\"ahler potential
\eqn\ktilde{{\cal K}^{(0)} = -\ln[-i(\tau -\bar \tau)] -
2\ln[\hat {\cal V}]
- \ln[-i \int_{Y_3} \Omega \wedge \bar \Omega]}
and {\gkp}
\eqn\dtau{\tilde D_{\tau} W = (\partial_\tau +
{\cal K}^{(0)}_\tau) W =
{1 \over \bar \tau - \tau} \int_{Y_3}
\bar G_{(3)} \wedge \Omega\ .}
The first correction term of \coralthree\ can be entirely
understood from
a Weyl-rescaling to the Einstein frame after the compactification.
More explicitly, reducing the ten-dimensional action in the
Einstein frame
leads to a non-canonically normalized
Einstein-Hilbert term in four dimensions (compare appendix)
\eqn\einhilb{-{1 \over 2 \kappa^2_{10}} \int d^{10} x \sqrt{-g}
(1 - {\zeta(3)\over 4} Q e^{-3/2 \phi_0}) R^{(10)}
\rightarrow -{1 \over 2 \kappa^2_{4}} \int d^{4} x \sqrt{-g}
(\hat {\cal V} + {\hat \xi \over 2}) R^{(4)}\ .}
Performing the Weyl-rescaling leads to the first correction term
to the potential.

The second term of \coralthree\ is more interesting.
Using \dtau\ and \super\ it is straightforward to verify that it
can be written as
\eqn\hcorr{-{3 \over 8}(\tau - \bar \tau)^2 {e^{{\cal K}^{(0)}} \over
\kappa^2_{10}} {\hat \xi \over \hat {\cal V}}
\int_{Y_3} H_{(3)} \wedge \Omega \int_{Y_3} H_{(3)} \wedge \bar
\Omega\ .}
This can be partly understood as follows. In the ten-dimensional
Type IIB action in
the string frame
there is a term

\eqn\hsquare{ {1 \over 4 \kappa^2_{10}} \int d^{10} x \sqrt{-g}
e^{-2\phi} {1 \over 3!} H_{MNP} H^{MNP}\ .}
Using \dilaton\ and performing the Weyl-rescaling
to the Einstein frame leads to a term
\eqn\hsquare{
{1 \over 4 \kappa^2_{10}} \int d^{10} x \sqrt{-g}
e^{-5/2\phi_0}(-{\zeta(3)\over8})
Q {1 \over 3!} H_{MNP} H^{MNP}\ .}
Reducing it on $Y_3$ leads to a term of the form \hcorr .
However, it does not reproduce the right factor and in addition from the
reduction of
\hsquare\ one would also expect further terms involving the (2,1)-forms
instead of the
(3,0)-form $\Omega$, c.f.\ the appendix of {\gkp}.
It is now natural to speculate that the higher derivative
corrections to the ten-dimensional Type IIB action provide the missing terms
in order
to give the result \hcorr\ in the reduction.
It would be interesting to
pursue this further and try to see if one might be able to put constraints
on possible higher derivative corrections involving two powers of $G$ and
three powers $R$
by demanding that they reproduce \hcorr .
For example one might find that
both $\sim (G^2 R^3 + c.c.)$- and
$\sim G \bar G R^3$-terms have to be present in ten dimensions.\foot{This
seems to be indicated by the fact that the correction terms in
{\hcorr} only depend on $H_{(3)}$ and not on $F_{(3)}$.}
However, a term $\sim G \bar G R^3$ is not to be expected
from an analysis using the linearized Type IIB
supersymmetry.\foot{M.H.~thanks
M.B.~Green and M.~Bianchi
for interesting discussions on this point.} On the other hand,
doubts have recently arisen that the superspace approach to the linearised
Type IIB theory
based on the scalar superfield of \ref\howest{P.S.~Howe and P.C.~West,
``The Complete ${\cal N}=2$, $D = 10$ Supergravity'', \np {238} {1984}
{181}.}
is capable of
capturing all the results of a string amplitude calculation performed in
{\pvwt}. Moreover, a
$G \bar G R^3$-term might be related to an $F^2 R^3$-term in
${\cal M}$-theory.
An argument for its presence has been put forward in
\ref\at{A.~A.~Tseytlin, ``$R^4$ Terms in $11$ Dimensions and Conformal
Anomaly of $(2,0)$ Theory'', \np {584} {2000}{233}, hep-th/0005072.}
based on the on-shell supergravity
superfield of eleven-dimensional supergravity \ref\cfbh{E.~J.~Cremmer and
S.~Ferrara, ``Formulation of Eleven-Dimensional Supergravity in
Superspace'',
\pl {91} {1980} {61}; L.~Brink and P.~Howe, ``Eleven-Dimensional
Supergravity
on the Mass-Shell in Superspace'', \pl {91} {1980} {384}.}.
Furthermore,
it seems to be necessary in order to explain the corrections to the
universal
hypermultiplet in ${\cal M}$-theory \ref\as{A.~Strominger, ``Loop
Corrections to the
Universal Hypermultiplet'', \pl {421} {1998} {139}, hep-th/9706195.}.
Finally, the corrections to the K\"ahler potential \kpot\ also lead to
modifications of the kinetic terms for $g_a$, the moduli stemming from
the four-form. Their occurrence indicates that there should be a term
$\sim F^2_{(5)} R^3$ in the ten-dimensional action.

We had seen in section 2 that the non-supersymmetric solutions
that we have
derived are
stable to leading order as they originate from the minimum of a positive
potential.
Here we see that for compactifications on manifolds with $\chi=0$
the solutions
are still stable as the corrections to the supergravity potential are
vanishing
in this case.
Furthermore, from formula {\coralthree} we see that the radial modulus is still not
stabilized
to order ${\cal O}(\alpha'^3)$, as we observe a runaway behaviour for it.
We expect that  additional corrections 
in $\alpha'$ indeed lead to a stabilization of the radial
modulus
because they will come with a different power of the volume.
If this was the case a cosmological
constant could be generated at higher order in $\a'$. It is tempting to
speculate that the sign of the cosmological constant might be
positive and that we could find de Sitter space as a space-time background.
It would certainly be wonderful if we could predict the
phenomenological correct relation between the supersymmetry
breaking scale and the cosmological constant
\ref\banks{T. Banks, ``Cosmological Breaking of Supersymmetry? or
Little Lambda Goes Back to the Future 2'', hep-th/0007146.}
in this way.

\bigskip

\noindent
{\bf Acknowledgments}

We would like to thank I.~Antoniadis, R.~D'Auria, M.~Berg, M.~Bianchi,
S.~Ferrara, S.~Giddings, M.~Green, S.~Gubser, M.~Luty, 
A.~Tseytlin, N.~Warner, E.~Witten and especially J.~Polchinski 
for interesting discussions.

K.Becker was supported by the US Department of Energy under grant
DE-FG03-92-ER40701 and in part by the University of Maryland's Center for
String Theory and Particle Theory. She wishes to thank the warm hospitality
from the Elementary Particle Theory Group at Maryland during the final
stages of this work. The work of M.~Becker was supported by NSF under
grant PHY-01-5-23911 and an Alfred Sloan Fellowship.
The work of M.~Haack was supported in part by
I.N.F.N., by the EC contract HPRN-CT-2000-00122, by the EC contract
HPRN-CT-2000-00148, by the INTAS contract 99-0-590 and by the MURST-COFIN
contract 2001-025492.
The work of J.~Louis was supported
by  the German Science Foundation (DFG),
the German--Israeli Foundation for Scientific Research (GIF),
the European RTN Program HPRN-CT-2000-00148 and
 the German Academic Exchange Service (DAAD).

\appendix{A}{Fixing the Constants in \prepot , \dilaton\ and \fourdil}

In this appendix we give some details of the derivation of the constants
appearing in \prepot\ and \dilaton\ and of the relation \fourdil.

The equation of motion for the dilaton
stemming from
\eqn\exo{S=-{1\over {2{\kappa}^2_{10}}}
\int d^{10} x \sqrt{-g^{(10)}}
e^{-2
\phi} (R+4(\partial \phi)^2 +(\alpha')^3 c_1 J_0 )\ ,}
to order ${\cal O}(\alpha'^3)$ is
\eqn\eom{
R+4 \nabla^2 \phi-4(\nabla \phi)^2 +\alpha'^3 c_1 J_0 =0\ .
}
We introduce complex coordinates on $Y_3$,
i.e.
\eqn\exii{
\xi^a = {1 \over \sqrt{2}} (y^{2a - 1} + i y^{2a})\qquad
{\rm for}\qquad  a=1, \ldots, 3\ .}
$J_0$ has the property of vanishing on Ricci flat K\" ahler
spaces. Therefore it does not contribute to \eom.

It has been shown in  {\grthree}, {\grtwo}
that up to four loops
the metric beta-function for the two-dimensional ${\cal N}=2$ non-linear
sigma model is given by\foot{Note that we use a convention for the
Ricci-tensor
that differs by a sign from the one used in {\grtwo}.}
\eqn\exiii{
\beta_{a {\bar b}}={1 \over 2 \pi}R_{a {\bar b}}+{1
\over 8 \pi} \zeta(3)
\nabla_a \nabla_{\bar b} Q\ ,
}
where we have again used $2 \pi \alpha' = 1$.
The explicit expression for $Q$ is
\eqn\q{
Q={1 \over 12 (2 \pi)^3 }\left({R_{IJ}}^{KL}{R_{KL}}^{MN}{R_{MN}}^{IJ}-2
R_{I\;\; J}^{\; \;K \;\; L}
R_{K\;\; L}^{\;\; M \;\; N}
R_{M\;\; N}^{\;\; I \;\; J}\right) .
}
For a six-dimensional manifold it is the Euler density,
i.e. $\int_{Y_3} d^6x \sqrt{g} Q = \chi$.

Demanding $\beta_{a {\bar b}}=0$ and using \eom\ and \exiii\
we see that the equation of motion is
satisfied to order ${\cal O}(\alpha'^3)$ if
\eqn\dil{\phi = \phi_0 + {\zeta(3) \over 16} Q\ ,}
where $\phi_0$ is a constant.

We now turn to the determination of the constant $\xi$ appearing in \prepot.
It is independent of the number of
K\"ahler moduli and therefore we consider the case of a single modulus here,
i.e.\ we make the following Ansatz for the metric in the string frame
\eqn\ansgar{ds^2 = \eta_{\mu \nu} dx^\mu dx^\nu + e^{2 u} \tilde g_{mn} dy^m
dy^n\ .}
We choose the volume of the background manifold
measured by $\tilde g$ to be $(2 \pi \alpha')^3=1$, such that
$\kappa_4^{-2}=\kappa_{10}^{-2}$. We
normalize the single (1,1)-form in such a way that we have ${\cal V} =
e^{6u} = v^3$, c.f. \volume . 

Fixing the constant in \prepot\ requires a reduction of \corr\ (augmented by
the term
\ddil), determining the kinetic term of $u$ and comparing
with the one obtained from \kcorr , i.e.
\eqn\kin{S = {1 \over \kappa_{4}^2} \int d^4 x \sqrt{-g} \Big( - (3 - 6 \xi
e^{-6u})
\partial_\mu u \partial^\mu u \Big) + \ldots\ .}
In order to derive \kin\ we have used that \kcorr\ is the K\"ahler
potential for the metric $G_{a \bar b}$ of the
K\"ahler moduli for both Type IIA and Type IIB on $Y_3$ and
the $z^a$ stand for the K\"ahler moduli $z^a = b^a - i v^a$.

In the reduction of $J_0$ we make use of the fact that it can be expressed
as {\fpss}\foot{See also \ref\grisaru{M.~T.~Grisaru and D.~Zanon,
``Sigma-Model Superstring
Corrections to the Einstein Hilbert Action'',
 \pl {177} {1986} {347}.}.}
\eqn\jzfpss{\eqalign{S=-{1\over {2{\kappa}^2_{10}}}
\int d^{10} x \sqrt{-g^{(10)}}
e^{-2\phi} \alpha'^3 c_2 \Big(12 Z - R S + 12 R_{MN} S^{MN} + {\rm
(Ricci)}^2 \Big) + \ldots \ ,}}
where we have introduced a new constant $c_2 = {\zeta(3) \over 3 \cdot 2^5}$
and used the
notation of \fpss , i.e.\ $Z=Z_{IJ} g^{IJ}$, $S=S_{IJ} g^{IJ}$ with
\eqn\zs{\eqalign{Z_{IJ} & = R_{IKLR}R_{JMN}\,\!\!^{R}
\Big( R^K\,\!\!_P\,\!\!^M\,\!\!_Q R^{NPLQ}
- {1 \over 2}R^{KN}\,\!\!_{PQ} R^{MLPQ}\Big)\ ,\cr
S_{IJ} & = -2 R_I\,\!\!^{MKL}R_J\,\!\!^P\,\!\!_K\,\!\!^Q R_{LPMQ} + {1 \over
2}R_I\,\!\!^{MKL}
R_{JMPQ} R_{KL}\,\!\!^{PQ} - R_I\,\!\!^K\,\!\!_J\,\!\!^L R_{KMNQ}
R_L\,\!\!^{MNQ}\ .}}
Furthermore, $S=12 (2 \pi)^3\, Q + {\rm (Ricci)}$, where $Q$ has been
defined in \q .

In order to evaluate \jzfpss\ for the metric \ansgar\ we need the
non-vanishing
components of the Riemann tensor. Using the conventions
\eqn\riemann{\eqalign{R^M\,\!\!_{NPQ} & = \partial_P \Gamma^M_{QN} -
\partial_Q \Gamma^M_{PN} + \Gamma^R_{QN} \Gamma^M_{PR}
- \Gamma^R_{PN} \Gamma^M_{QR} \cr
\Gamma^M_{NP} & = {1 \over 2} g^{MQ} \Big( \partial_N g_{PQ} + \partial_P
g_{QN}
-\partial_Q g_{NP} \Big)\ ,}}
we find
\eqn\ntriem{\eqalign{R^m\,\!\!_{\mu n \nu} & = -\delta^m_n (\partial_\mu u
\partial_\nu u
+ \partial_\mu \partial_\nu u)\ , \cr
R^\mu\,\!\!_{m \nu n} & = - g_{mn} (\partial_\nu u \partial^\mu u
+ \partial_\nu \partial^\mu u)\ , \cr
R^k\,\!\!_{m n p} & = \tilde R^k\,\!\!_{m n p} + \partial_\mu u \partial^\mu
u
(\delta^k_p g_{mn} - \delta^k_n g_{pm})}}
with all other components, not related to these by symmetry, vanishing. Thus
the
non-trivial components of the Ricci-tensor are\foot{Here
we skip the contribution of the background metric. According
to {\exiii} and in complex coordinates it is proportional to
$\nabla_a \nabla_{\bar b} \tilde Q$ and does not
contribute in the reduction.}
\eqn\ntricci{\eqalign{R_{\mu \nu} & = - 6 (\partial_\mu u \partial_\nu u
+ \partial_\mu \partial_\nu u)\ , \cr
R_{mn} & = - g_{mn} (6\partial_\mu u \partial^\mu u
+ \partial_\mu \partial^\mu u)\ ,}}
whereas the Ricci scalar is given by
\eqn\riccis{R = -42 \partial_\mu u \partial^\mu u - 12 \partial_\mu
\partial^\mu u\ .}

Now we are in a position to discuss the different sources for contributions
to the
kinetic term of $u$ stemming from \jzfpss . {}From \ntricci\ and \riccis\ it
is clear that
the terms $\sim RS$ and $\sim R_{mn} S^{mn}$ contribute, whereas
the $({\rm Ricci})^2$-terms and the one $\sim R_{\mu\nu} S^{\mu\nu}$ do not.
More
involved is the discussion of the $Z$-term. {}From its definition in \zs\ we
see that
the only contributions come from taking all four Riemann tensors with
internal
indices only, i.e.\ those given in the last line of \ntriem . To second
order in
derivatives of $u$ we get
\eqn\z{Z =  e^{-6 u} \partial_\mu u \partial^\mu u \Big( 12 (2 \pi)^3 \tilde
Q
+ \tilde R^i\,\!\!_{qlr} \left(2 \tilde R_{ipn}\,\!\!^r \tilde R^{nplq}
- \tilde R_{inp}\,\!\!^r \tilde R^{nlpq} - \tilde R_{ipn}\,\!\!^r
\tilde R^{qnlp}\right) \Big)\ ,}
where $\tilde Q$ is as in \q\ but evaluated with the background metric
$\tilde g_{mn}$.
The second cubic polynomial in the Riemann tensor is different from $\tilde
Q$
for a general manifold. On a K\"ahler manifold, however, it is possible to
show
that it is indeed given by $12 (2 \pi)^3 \tilde Q$. To see this we again
introduce
complex coordinates on $Y_3$.
Using the fact that the only non-trivial (independent) component of the
Riemann tensor on a K\"ahler manifold, $\tilde R^a\,\!\!_{bc\bar d}$, has
the additional symmetry
\eqn\adsym{\tilde R^a\,\!\!_{bc\bar d} = \tilde R^a\,\!\!_{cb\bar d}\ ,}
we derive for K\"ahler manifolds
\eqn\qkaehler{\eqalign{12 (2 \pi)^3 \tilde Q & = 4 \left(
\tilde R_{a \bar b}\,\!^{c \bar d}
\tilde R_{c \bar d}\,\!^{e \bar f} \tilde R_{e \bar f}\,\!^{a \bar b} -
\tilde R_a\,\!^c\,\!_b\,\!^d
\tilde R_c\,\!^e\,\!_d\,\!^f \tilde R_e\,\!^a\,\!_f\,\!^b \right) \cr
& = \tilde R^i\,\!\!_{qlr} \left(2 \tilde R_{ipn}\,\!\!^r \tilde R^{nplq}
- \tilde R_{inp}\,\!\!^r \tilde R^{nlpq} - \tilde R_{ipn}\,\!\!^r
\tilde R^{qnlp}\right)\ .}}
Thus on a K\"ahler manifold \z\ simplifies and becomes
\eqn\zkaehler{Z =   24 (2 \pi)^3 e^{-6 u} \partial_\mu u \partial^\mu u
\tilde Q\ .}

Let us now perform the reduction of \corr\ with the metric Ansatz \ansgar\
and
the dilaton $\phi = \phi_0 + c Q$, where the constant $c$ has been
determined in \dil\ to be $c= \zeta(3) 2^{-4}$. As we have already
mentioned in the main text we include a term
\eqn\dddil{-{a\over {2{\kappa}^2_{10}}}
\int d^{10} x \sqrt{-g^{(10)}} e^{-2\phi} \alpha'^3 c_2 (\nabla^2 \phi) S\ .}
The constant $a$ can be determined as follows. It has been argued in
\ref\pz{Q.H.~Park and D.~Zanon,
``More on Sigma Model Beta Functions and Low-Energy Effective Actions'',
\physrev {35} {1987} {4038}.} that the Ricci-terms in {\jzfpss} actually
appear in the combination $R_{MN} + 2 \nabla_M \nabla_N \phi$.\foot{Note
that
our definition of the dilaton differs from {\pz} by a factor of $-2$ but
also the
definition of our Ricci-tensor is different by a factor of $-1$.}
Inserting this
into {\jzfpss} we get the additional terms
\eqn\add{S=-{1\over {2{\kappa}^2_{10}}} \int d^{10} x \sqrt{-g^{(10)}}
e^{-2\phi} \alpha'^3 c_2 \Big( - 2 (\nabla^2 \phi) S
+ 24 (\nabla_M \nabla_N \phi) S^{MN} \Big) + \ldots\ .}
Whereas the second one does not contribute in the reduction,
the first one is exactly of the form
{\dddil} with a constant $a=-2$.

Using this we derive to order ${\cal O}(\alpha'^3)$
\eqn\sred{\eqalign{S = -{1 \over 2 \kappa_{10}^2} \int d^{10} x \sqrt{-g}
e^{- 2\phi_0}
\Big[ & (1 - 2 c Q) \Big(R^{(4)} - 42 \partial_\mu u \partial^\mu
u
- 12 \partial_\mu \partial^\mu u
+ 4 \partial_\mu \phi_0 \partial^\mu \phi_0 \Big) \cr
& - 48 c Q \partial_\mu \phi_0 \partial^\mu u
-48 c_2 Q \partial_\mu \phi_0 \partial^\mu \phi_0
\cr
& + 12 c_2 Q (-R^{(4)} - 6 \partial_\mu u \partial^\mu
u) \Big]\ .}}
Integrating over the internal coordinates and
using $6 c_2 = {1 \over 16} \zeta(3) = c$ we
arrive at
\eqn\sredfour{\eqalign{S = -{1 \over 2 \kappa_{4}^2} \int d^4 x \sqrt{-g}
e^{- 2\phi_0}
& \Big[ (e^{6 u} - 4 c  \chi) R^{(4)} + (e^{6 u} - 2 c \chi) ( - 42
\partial_\mu u \partial^\mu u
- 12 \partial_\mu \partial^\mu u)  \cr
& + \left( e^{6 u} - 4 c \chi \right) 4 \partial_\mu \phi_0
\partial^\mu \phi_0
- 48 c \chi \partial_\mu \phi_0 \partial^\mu u
- 12 c \chi \partial_\mu u \partial^\mu u \Big]\ .}}
{}From the prefactor of $R^{(4)}$ we can read off the four-dimensional
dilaton
\eqn\fdil{e^{-2\phi_4} = e^{-2\phi_{0}} \Big( e^{6u} - 4c \chi \Big)\ . }
Performing a partial integration in the last term of the first row of
\sredfour\  it
is straightforward to verify that \sredfour\ can be expressed
as\foot{Without the
term {\dddil} one would get unallowed cross terms
involving the four-dimensional
dilaton and the radial mode $u$.}
\eqn\sredfdil{\eqalign{S=-{1 \over \kappa_{4}^2} \int d^4 x \sqrt{-g} e^{-
2\phi_4}
\Big({1 \over 2} R^{(4)} + 2 \partial_\mu \phi_4 \partial^\mu \phi_4
- [3 + 48 c \chi e^{-6u}] \partial_\mu u \partial^\mu u \Big)\ .}}
A Weyl-rescaling to the four-dimensional Einstein frame does not
alter the
coefficient of the kinetic term for $u$. We can therefore compare it with
\kin\ and find
\eqn\kincomp{\xi = -8 c \chi = - {\zeta(3) \chi \over 2}\ .}
Inserting this into {\fdil} leads to {\fourdil}.

\listrefs

\end